\documentclass[a4paper,onecolumn,12pt]{article}
\usepackage{geometry}
\usepackage[natbib=true,backend=biber,sorting=nyt,style=apa,sortcites=true]{biblatex}
\usepackage[utf8]{inputenc}
\usepackage{authblk}
\usepackage{graphicx}
\usepackage[font=small]{caption}
\usepackage{eurosym}
\usepackage{hyperref}
\usepackage{xcolor}
\hypersetup{
    colorlinks,
    linkcolor={red!60!black},
    citecolor={blue!60!black},
    urlcolor={blue!90!black}
}

\title{Computational Desire Line Analysis of Cyclists\\on the Dybbølsbro Intersection in Copenhagen}
\author[a]{Simon Martin Breum}
\author[a]{Bojan Kostic}
\author[a,b,c]{Michael Szell\thanks{Corresponding author. Email: \href{mailto:misz@itu.dk}{misz@itu.dk}}}

\affil[a]{NEtwoRks, Data, and Society (NERDS), IT University of Copenhagen, 2300 Copenhagen, Denmark}
\affil[b]{ISI Foundation, 10126 Turin, Italy}
\affil[c]{Complexity Science Hub Vienna, 1080 Vienna, Austria}

\date{}

\bibliography{main.bib}

\begin{document}

\maketitle

\vspace{-1.4cm}
\begin{abstract}
\noindent
Contemporary street design prioritizes vehicular traffic flow and assumes compliant road users. However, actual human behavior is typically neglected, especially of cyclists, leading to streets with inadequate wayfinding and protection from vehicular traffic. To improve planning, here we develop a computational method to detect cyclist trajectories from video recordings and apply it to the Dybbølsbro intersection in Copenhagen, Denmark. In one hour of footage we find hundreds of trajectories that contradict the design, explainable by the desire for straightforward, uninterrupted travel largely not provided by the intersection. This neglect and the prioritization of vehicular traffic highlight opportunities for improving Danish intersection design.\\[10px]
\end{abstract}

{\vskip 10mm}

\section{Questions}
Safe and functional cycling infrastructure is necessary to support the uptake of cycling in cities \citep{winters2017policies}. Especially street intersections are important conflict points where cars and bicycles meet, causing a large fraction of road deaths and injuries \citep{gotschi2018towards,dozza2014introducing,ling2019cmv,bahrololoom2020mis}, and must therefore be planned with human behavior in mind. The intersection at Dybbølsbro, Copenhagen, is a notorious example which has been criticized for confusing cyclists due to its difficulty to navigate, and is currently scheduled for a second major redesign \citep{hunter2021,therkildsen2021,wsp}. To understand to which extent intersection designs are adequate for cyclists, some studies have begun tracing and recording cyclist trajectories and behavior \citep{colville2013bicycle,te2014choreografie,lindrule2021,nabavi_niaki_is_2019,casello_enhancing_2017}. However, these methods are manual, therefore costly and not scalable.

Here we first ask: How can we use computational methods to automatize the analysis of cyclist trajectories? Focusing on Dybbølsbro, we then ask: How much do cyclist trajectories deviate from the design's intended paths and why? Finally: What are the implications for the design of the Dybbølsbro intersection and of Danish intersections in general?

\begin{figure}[t]
\centering
\includegraphics[width=0.9\textwidth]{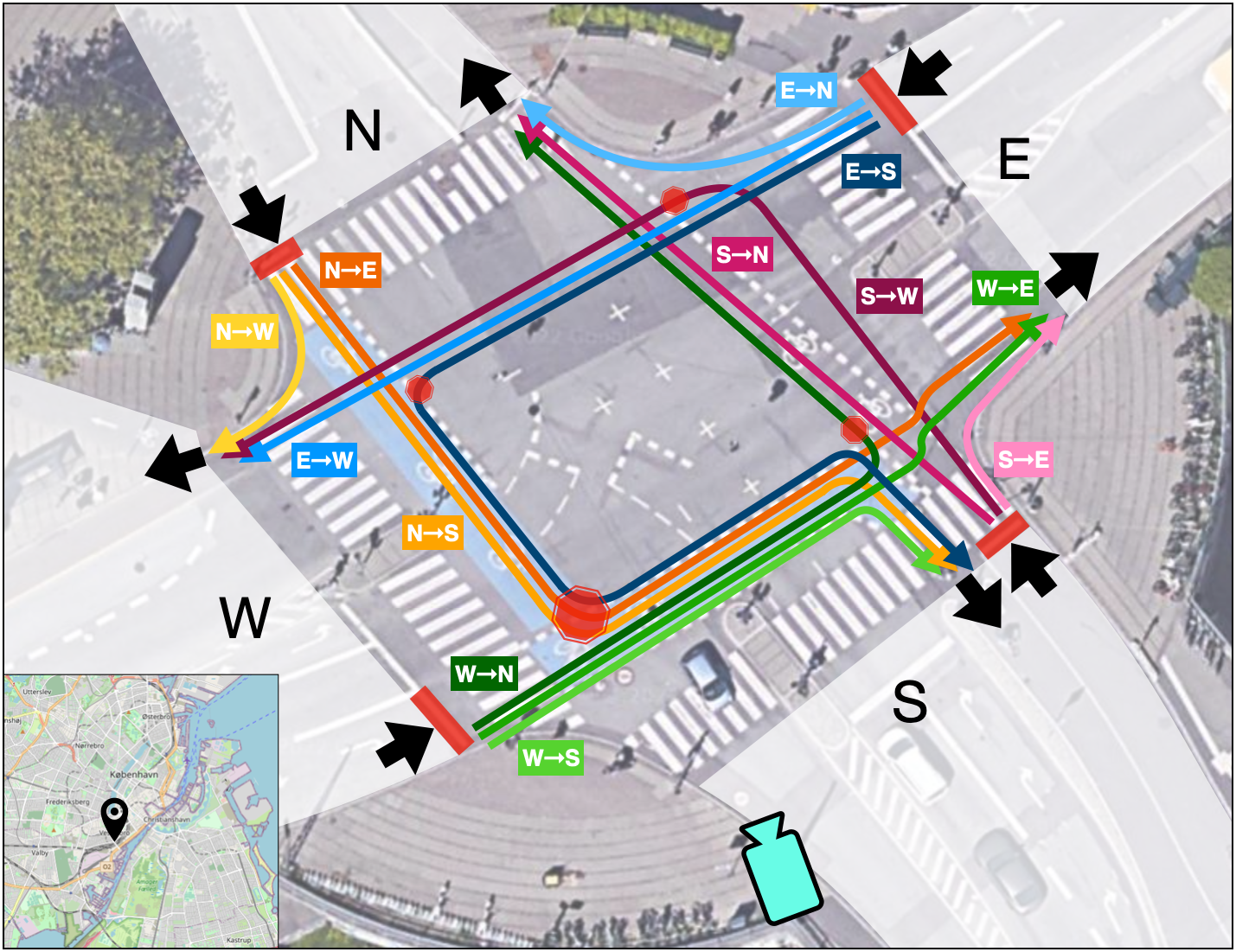}
\caption{\textbf{The study area is the Dybbølsbro intersection in Copenhagen (June 2021).} There are 12 possible designed source-destination paths (colored arrows) between the four sides N, W, S, E, connecting 8 legal entry and exit points for cyclists (black arrows). Before entering the intersection, cyclists on every side have to wait behind a signal line at red (red lines). Cyclists going from N to S (N$\rightarrow$S) must in practice take an additional stop (stop symbol) due to the switch to a bidirectional cycle track on the S side. Left turning cyclists are technically allowed to turn left straightaway against red like vehicular traffic \citep{larsen2017inhabiting}, but this is uncommon practice; in practice, cyclists must include one additional stop at a red traffic light in the corner when traveling S$\rightarrow$W, N$\rightarrow$E, W$\rightarrow$N and two stops when traveling E$\rightarrow$S. The camera symbol depicts the camera mounting at 10m height. Map data: \copyright 2022 Google / Aerodata International Surveys, Maxar Technologies, and \copyright OpenStreetMap contributors.\label{fig:dybbelsbro}}
\end{figure}  

\section{Methods}
All data and code to reproduce our findings are available at: \href{https://github.com/SimonBreum/desirelines}{github.com/SimonBreum/desirelines}  Our starting point is a set of 11,553 cyclist trajectories, which had been extracted via a custom-trained YOLO model \citep{Redmon2018} from a high-resolution 1h video from 2021-06-09 07:00-08:00 (Wednesday) of the Dybbølsbro intersection, see Fig.~\ref{fig:dybbelsbro}. This intersection has been redesigned in 2019 with a bidirectional bicycle track on the south side (S), which has made it difficult for cyclists to navigate due to the need to switch sides when coming from north (N) \citep{wsp}. See Fig.~\ref{fig:dybbelsbro} for all possible designed paths (simplifying one additional street in the northwest). Apart from the unconventional N$\rightarrow$S path, the left turns S$\rightarrow$N, N$\rightarrow$E, W$\rightarrow$N require one additional stop, and the left turn E$\rightarrow$S requires two, due to general Danish intersection design (the ``Copenhagen left'')\citep{larsen2017inhabiting}.

\begin{figure}[t]
\centering
\includegraphics[width=0.99\textwidth]{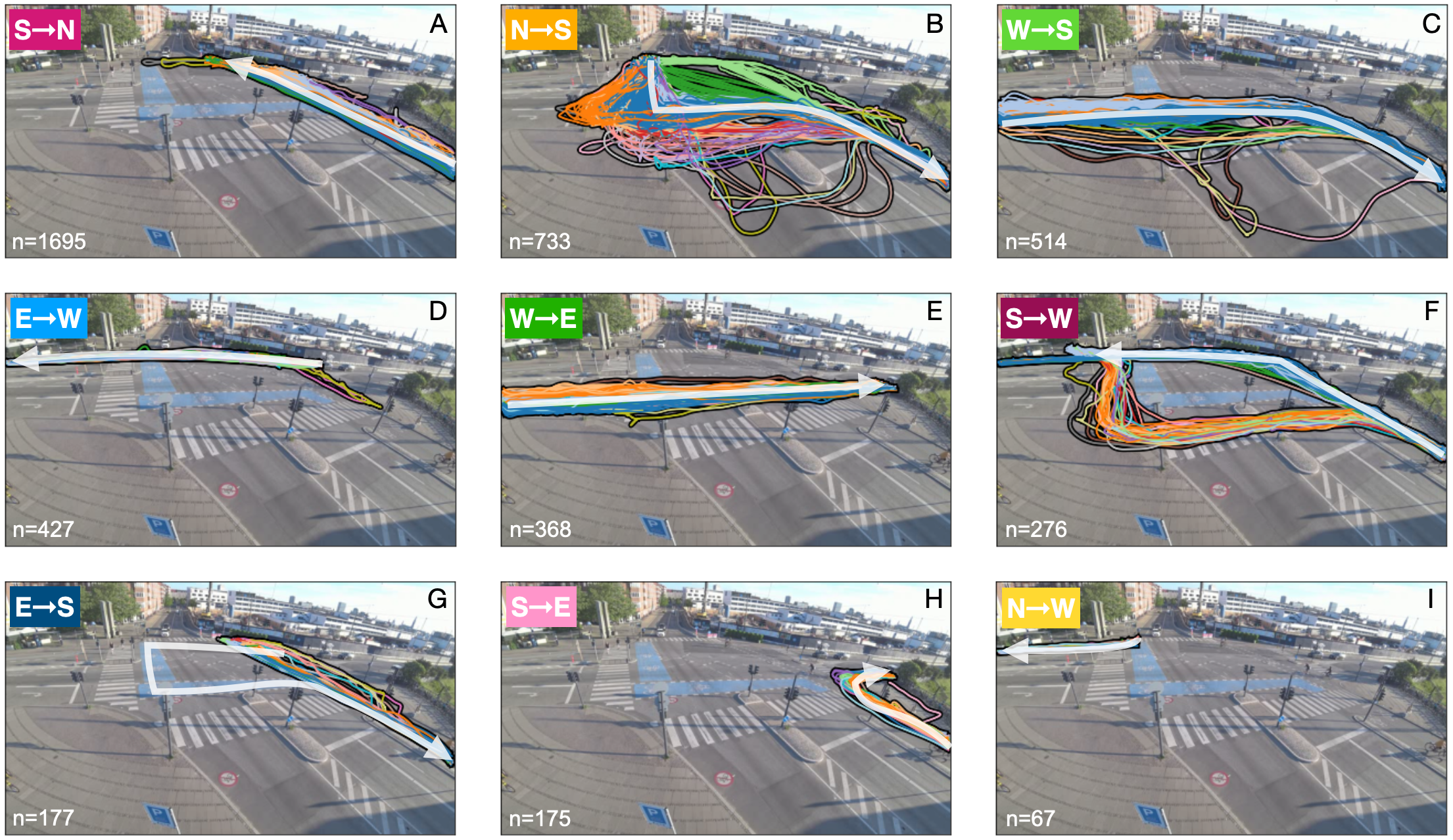}
\caption{\textbf{Traversal of the 12 source-destination (SD) pairs by the cyclists, which yields 12 SD-clusters of trajectories.} White arrows denote the intended paths. Many trajectories deviate substantially from the intended paths. Colors within each SD-cluster depict different path clusters. For the three SD pairs N$\rightarrow$E, W$\rightarrow$N and E$\rightarrow$N there are not enough trajectories to detect an SD-cluster.\label{fig:sdclusters}}
\end{figure}

We applied DBSCAN to source-destination pairs with $\varepsilon=8$ pixels (at a $640\times 360$ resolution) and $\mathrm{minPts}=25$, which yielded 4888 trajectories distributed among 16 source-destination (SD) clusters. We discarded the remaining 6665 trajectories which are mostly broken trajectories, for example due to occlusion by traffic signs or vehicles. After manual inspection we merged two pairs of SD-clusters that each had the same source and destination. We also discarded three other clusters of broken trajectories. In total this yielded 9 SD-clusters with 4432 trajectories, see Fig.~\ref{fig:sdclusters}, matching the 12 possible designed paths from Fig.~\ref{fig:dybbelsbro} except for N$\rightarrow$E, W$\rightarrow$N, and E$\rightarrow$N where not enough trajectories were found.

\begin{figure}[t]
\centering
\includegraphics[width=0.99\textwidth]{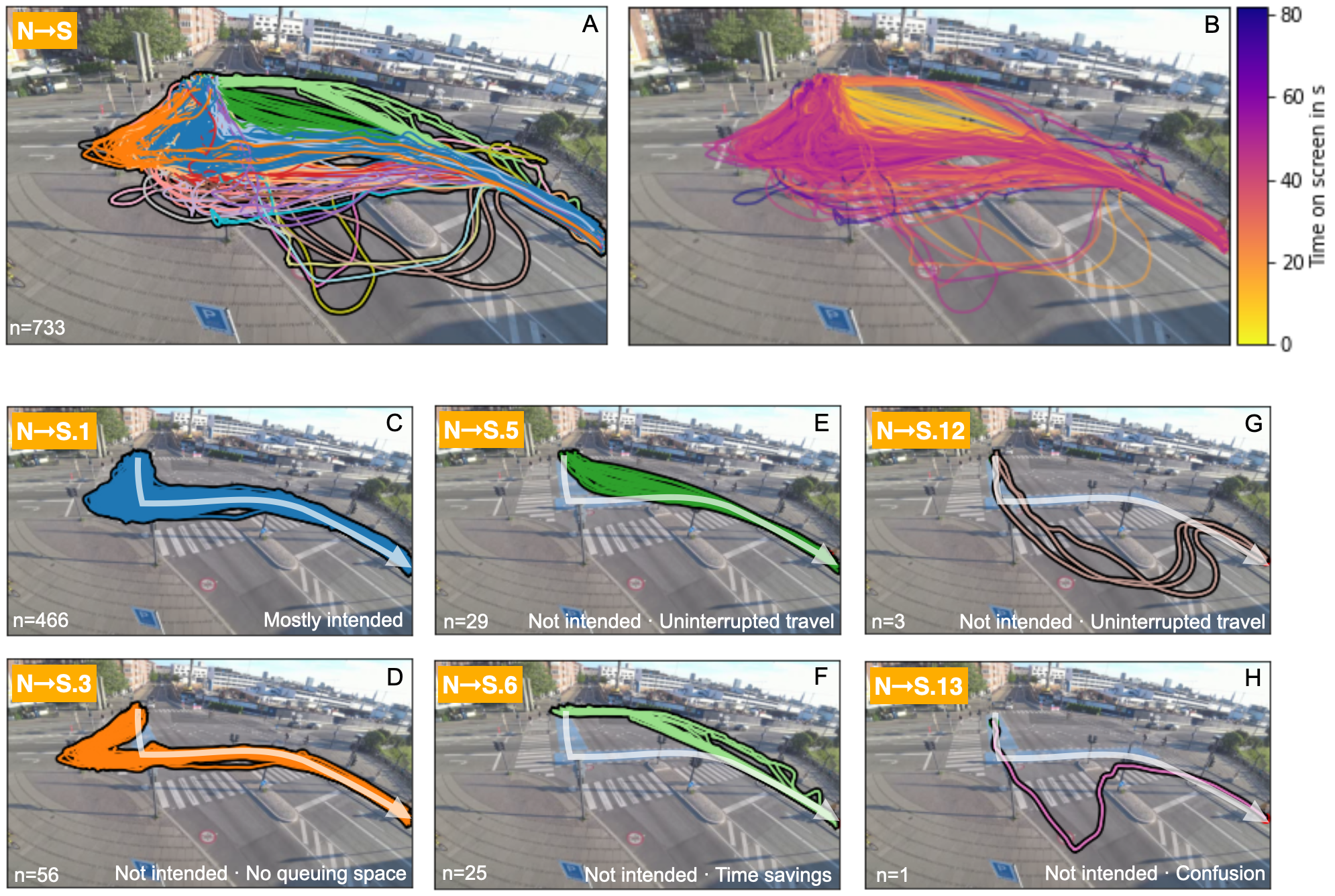}
\caption{\textbf{Investigation of SD-cluster N$\rightarrow$S and some of its path-clusters.} A) SD-cluster N$\rightarrow$S and its 733 trajectories. B) Time on screen demonstrates that crossing the intersection diagonally, while illegal and not intended by the planners, provides substantially shorter crossing time (13s on average) than following the intended path (43s on average). Crossing via the NE corner is also faster (32s on average) than the intended path. C) Path-cluster N$\rightarrow$S.1 shows mostly intended behavior (466 trajectories). D) Path-cluster N$\rightarrow$S.3 shows not intended behavior because cyclists move onto the cross-walk, presumably due to lack of queuing space (56 trajectories). E) Path-cluster N$\rightarrow$S.5 shows not intended behavior, crossing diagonally (29 trajectories). F) Path-cluster N$\rightarrow$S.6 shows not intended behavior, crossing via the NE corner instead of the SW corner (25 trajectories). G-H) Path-clusters N$\rightarrow$S.13 and N$\rightarrow$S.12 show not intended behavior, entering street space that is intended for cars only, possibly due to uninterrupted travel or confusion (3 and 1 trajectories, respectively).\label{fig:pathclusters}}
\end{figure}  

To each of the SD-clusters we applied dynamic time warping \citep{berndt1994using}, generating 20 additional path-clusters respectively, denoted by different trajectory colors in Fig.~\ref{fig:sdclusters}. Finally, we contrasted these path-clusters with the designed paths to study how cyclists are actually moving from each source to each destination versus how it was intended by the planners.

\section{Findings}
We found that at least 11\% (495 out of 4432) trajectories are not following the designed paths. The effect is particularly strong in two specific SD-clusters:
\begin{itemize}
    \item \textbf{SD-cluster N$\rightarrow$S (Fig.~\ref{fig:pathclusters}).} Path-cluster N$\rightarrow$S.1 (Fig.~\ref{fig:pathclusters}C): Only 466 out of 733 cyclists follow mostly intended behavior, implying a mismatch between design and reality of at least 36\%. Path-cluster N$\rightarrow$S.3 (Fig.~\ref{fig:pathclusters}D): Due to lack of queuing space, many cyclists cannot wait in front of the pedestrian crossing but are forced to enter it. Path-cluster N$\rightarrow$S.5 (Fig.~\ref{fig:pathclusters}F): 29 cyclists crossed the intersection diagonally. Path-cluster N$\rightarrow$S.6 (Fig.~\ref{fig:pathclusters}E): 25 cyclists crossed via the NE corner instead of the SW corner. Analysis of trajectory durations reveals the likely cause (Fig.~\ref{fig:pathclusters}B): On average, diagonally crossing cyclists spend only 13s, and cyclists crossing via the NE corner spend 32s. Contrast these values to 43s, which is the time spent by cyclists who follow the designed path with the additional stop. Further path-clusters (Fig.~\ref{fig:pathclusters}G,H): Uninterrupted, fast travel (3 cyclists), and confusion (1 cyclist).
    \item \textbf{SD-cluster E$\rightarrow$S.} Here we found 0 out of 177 empirical trajectories following the intended path, implying a mismatch between design and reality of 100\%, explainable with the two additional stops that are considerably more convoluted than the direct path, Fig.~\ref{fig:sdclusters}G. To double-check, we selected from all 11,553 trajectories those going E$\rightarrow$S irrespective of clustering, and found 9 out of 518 taking the two additional stops, lowering the mismatch to 98\%.
\end{itemize}

Apart from SD-cluster-specific issues, we also counted 12 trajectories from several SD-clusters that enter the wrong -- vehicles-only -- side on the southern street like in Fig.~\ref{fig:pathclusters}G,H. Although the fraction of these trajectories is small, they represent potentially dangerous situations where cyclists are traversing three vehicular lanes.

We have shown that our mostly automated method can well support the behavioral analysis of a large number of cyclists, and it has quantified a non-negligible number of at least 495 not intended, potentially life-threatening trajectories -- all happening in just one hour. It is an open question whether our method can be generalized and fully automatized, and how the quality of analysis compares to manual methods. In the future, every step of our computational pipeline should be scrutinized to ensure high trajectory quality. In particular, bias could have been introduced by lost trajectories from occlusion or tracking errors in specific parts of the study area. In any case, we expect our method to scale better and to be less costly.

For the upcoming re-design of the Dybbølsbro intersection, consultants have considered traffic counts from video analysis and qualitative assessment of behavior, but without quantifying desire lines \citep{wsp}. A repeated evaluation with our method after implementation could provide an assessment of the re-design's success rate, and whether a more profound analysis or re-design is called for. Our results confirm the intentions of the re-design \citep{wsp} that intersection complexity should be lowered and the momentum and smooth wayfinding for cyclists should be respected, as also found in previous research from Spain \citep{lindrule2021}, the Netherlands \citep{hahn_collaboration_2021}, Canada \citep{nabavi_niaki_is_2019}, and Denmark \citep{colville2013bicycle}. However, the mixing of a bidirectional lane with unidirectional lanes remains particularly problematic \citep{lindrule2021}, as does the lack of queuing space and protection for cyclists \citep{planamsterdam,nacto2014ubd}.

The underlying issue is the prioritization of vehicular traffic flow in Danish street design, which persists despite successful efforts at improving cycling \citep{nielsen2013urban,colvilleandersen2017asc,colville2013bicycle,szell2018cqv}. As we have shown, this priority leads to additional interruptions for cyclists, forcing traffic violations and competition with pedestrian space. Due to the skewed threat posed by vehicular traffic \citep{verkade2019,klanjcic2021iuf}, such violations are most hazardous to the cyclists themselves. Following research and best practices in road safety \citep{aldred2018cir,marshall2019wcw,hartmann2020hoa,branioncalles2020ccr,nieuwenhuijsen2020utp,world2022walking}, the acceptable level of vehicular traffic flow should be well justified. If this level is above zero, known effective solutions can include transformation of vehicular space into more queuing space, drastic speed reductions with possible removal of traffic lights, or similar improvements \citep{hahn_collaboration_2021,nacto2014ubd,planamsterdam}. However, such considerations are not part of the upcoming re-design where car traffic cannot be obstructed \citep{wsp}. It is an open research question why that is the case \citep{mattioli2020political,gossling2020cities}, given the projected increase of cycling \citep{wsp}, and that the private car is the most hazardous \citep{cantuaria2021residential,klanjcic2021iuf}, unsustainable \citep{banister2005unsustainable}, and societally uneconomic \citep{gossling2019sca} mode of urban transport.

\section*{Acknowledgments}
This study was supported by the Danish Ministry of Transport. We thank COWI for video capture, the NERDS group, especially Ane Rahbek Vierø and Anastassia Vybornova, for helpful discussions, and ITU's HPC for computational support.


\printbibliography

@article{colville2013bicycle,
  title={The Bicycle Choreography of an Urban Intersection},
  author={Colville-Andersen, Mikael and Madruga, Pedro and Kujanp{\"a}{\"a}, Risto and Maddox, Kristen},
  journal={Desire Lines \& Behaviour of Copenhagen Bicycle Users. Copenhagenize Design Co. Frederiksberg, Denmark},
  year={2013}
}

@article{nabavi_niaki_is_2019,
	title = {Is that move safe? {Case} study of cyclist movements at intersections with cycling discontinuities},
	volume = {131},
	issn = {00014575},
	shorttitle = {Is that move safe?},
	url = {https://linkinghub.elsevier.com/retrieve/pii/S0001457518306092},
	doi = {10.1016/j.aap.2019.07.006},
	language = {en},
	urldate = {2022-09-10},
	journal = {Accident Analysis \& Prevention},
	author = {Nabavi Niaki, Matin S. and Saunier, Nicolas and Miranda-Moreno, Luis F.},
	month = oct,
	year = {2019},
	pages = {239--247},
}

@article{hahn_collaboration_2021,
	title = {Collaboration, experimentation, continuous improvement: {Exploring} an iterative way of working in the {Municipality} of {Amsterdam}’s {Bicycle} {Program}},
	volume = {9},
	issn = {25901982},
	shorttitle = {Collaboration, experimentation, continuous improvement},
	url = {https://linkinghub.elsevier.com/retrieve/pii/S2590198220302001},
	doi = {10.1016/j.trip.2020.100289},
	language = {en},
	urldate = {2022-09-12},
	journal = {Transportation Research Interdisciplinary Perspectives},
	author = {Hahn, Trey and {t}e Brömmelstroet, Marco},
	month = mar,
	year = {2021},
	pages = {100289}
}

@techreport{planamsterdam,
     title = {Plan Amsterdam - Giving way to cyclists},
     year = {2018},
     author = {{Gemeente Amsterdam}},
     url={https://handshakecycling.eu/resources/plan-amsterdam-giving-way-cyclists}
}

@techreport{wsp,
     title = {DOBBELTRETTET CYKELSTI I KRYDSET INGERSLEVSGADE-DYBBØLSBRO, DISPOSITIONSFORSLAG},
     year = {2021},
     author = {{WSP DANMARK A/S}},
     url={https://vesterbrolokaludvalg.kk.dk/sites/default/files/2022-05/Dispositionsforslag_No.pdf}
}

@incollection{larsen2017inhabiting,
  title={Inhabiting infrastructures: the case of cycling in {C}openhagen},
  author={Larsen, Jonas and Funk, Oskar},
  booktitle={Experiencing Networked Urban Mobilities},
  pages={129--134},
  year={2017},
  publisher={Routledge}
}

@article{nielsen2013urban,
  title={Urban planning practices for bikeable cities--the case of {C}openhagen},
  author={Nielsen, Thomas A Sick and Skov-Petersen, Hans and Agervig Carstensen, Trine},
  journal={Urban Research \& Practice},
  volume={6},
  number={1},
  pages={110--115},
  year={2013},
  publisher={Taylor \& Francis},
  doi={10.1080/17535069.2013.765108}
}

@inproceedings{te2014choreografie,
  title={De choreografie van een kruispunt: Naar een gebruiksgeorienteerde ontwerplogica voor kruispunten},
  author={{t}e Br{\"o}mmelstroet, Marco},
  booktitle={Colloquium Vervoersplanologisch Speurwerk},
  pages={1--15},
  year={2014}
}

@inproceedings{berndt1994using,
  title={Using dynamic time warping to find patterns in time series.},
  author={Berndt, Donald J and Clifford, James},
  booktitle={KDD workshop},
  volume={10},
  number={16},
  pages={359--370},
  year={1994},
  organization={Seattle, WA, USA:}
}

@article{hunter2021,
  title={Local Round-Up: City planners rethink problematic Fisketorvet junction – again!},
  author={Hunter, Lena},
  year={2021},
  journal={CPH Post},
  url={https://cphpost.dk/?p=123482}
}

@article{therkildsen2021,
  title={Omstridt og berygtet lyskryds får løsning, der aldrig før er set i Danmark},
  author={Therkildsen, C.S.N.},
  year={2021},
  journal={TV 2 Lorry},
  url={https://www.tv2lorry.dk/koebenhavn/omstridt-og-berygtet-lyskryds-faar-loesning-der-aldrig-foer-er-set-i-danmark}
}

@article{dozza2014introducing,
  title={Introducing naturalistic cycling data: What factors influence bicyclists’ safety in the real world?},
  author={Dozza, Marco and Werneke, Julia},
  journal={Transportation research part F: traffic psychology and behaviour},
  volume={24},
  pages={83--91},
  year={2014},
  publisher={Elsevier},
  doi={10.1016/j.trf.2014.04.001}
}

@article{gotschi2018towards,
  title={Towards a comprehensive safety evaluation of cycling infrastructure including objective and subjective measures},
  author={G{\"o}tschi, Thomas and Castro, Alberto and Deforth, Manja and Miranda-Moreno, Luis and Zangenehpour, Sohail},
  journal={Journal of Transport \& Health},
  volume={8},
  pages={44--54},
  year={2018},
  publisher={Elsevier},
  doi={10.1016/j.jth.2017.12.003}
}

@article{winters2017policies,
  title={Policies to promote active travel: evidence from reviews of the literature},
  author={Winters, Meghan and Buehler, Ralph and G{\"o}tschi, Thomas},
  journal={Current environmental health reports},
  volume={4},
  number={3},
  pages={278--285},
  year={2017},
  publisher={Springer},
  doi={10.1007/s40572-017-0148-x}
}

@article{casello_enhancing_2017,
	title = {Enhancing {Cycling} {Safety} at {Signalized} {Intersections}: {Analysis} of {Observed} {Behavior}},
	volume = {2662},
	issn = {0361-1981, 2169-4052},
	shorttitle = {Enhancing {Cycling} {Safety} at {Signalized} {Intersections}},
	url = {http://journals.sagepub.com/doi/10.3141/2662-07},
	doi = {10.3141/2662-07},
	language = {en},
	number = {1},
	urldate = {2022-09-10},
	journal = {Transportation Research Record: Journal of the Transportation Research Board},
	author = {Casello, Jeffrey M. and Fraser, Adam and Mereu, Alex and Fard, Pedram},
	month = jan,
	year = {2017},
	pages = {59--66},
}

@article{lindrule2021,
	title = {Rule compliance and desire lines in {Barcelona}’s cycling network},
	volume = {13},
	issn = {1942-7867, 1942-7875},
	url = {https://www.tandfonline.com/doi/full/10.1080/19427867.2020.1803542},
	doi = {10.1080/19427867.2020.1803542},
	language = {en},
	number = {10},
	urldate = {2022-09-10},
	journal = {Transportation Letters},
	author = {Lind, Adam and Honey-Rosés, Jordi and Corbera, Esteve},
	month = nov,
	year = {2021},
	pages = {728--737},
}

@article{klanjcic2021iuf,
  title={Identifying urban features for vulnerable road user safety in {E}urope},
  author={Klanj\v{c}i{\'c}, Marina and Gauvin, Laetitia and Tizzoni, Michele and Szell, Michael},
  journal={EPJ Data Science},
  year={2022},
  doi = {10.1140/epjds/s13688-022-00339-5}
}

@article{verkade2019,
title={Het grootste taboe in het verkeer: we kunnen elkaar doodrijden.},
author={Verkade, T. and {t}e Brömmelstroet, M.},
journal={De Correspondent.},
year={2019},
url={https://decorrespondent.nl/9156/het-grootste-taboe-in-het-verkeer-we-kunnen-elkaar-doodrijden/3738007565292-e9168246}

}

@article{cantuaria2021residential,
  title={Residential exposure to transportation noise in {D}enmark and incidence of dementia: national cohort study},
  author={Cantuaria, Manuella Lech and Waldorff, Frans Boch and Wermuth, Lene and Pedersen, Ellen Raben and Poulsen, Aslak Harbo and Thacher, Jesse Daniel and Raaschou-Nielsen, Ole and Ketzel, Matthias and Khan, Jibran and Valencia, Victor H and others},
  journal={bmj},
  volume={374},
  year={2021},
  publisher={British Medical Journal Publishing Group},
  doi={10.1136/bmj.n1954}
}

@book{banister2005unsustainable,
  title={Unsustainable transport: city transport in the new century},
  author={Banister, David},
  year={2005},
  publisher={Routledge},
  doi={10.4324/9780203003886}
}

@article{mattioli2020political,
  title={The political economy of car dependence: A systems of provision approach},
  author={Mattioli, Giulio and Roberts, Cameron and Steinberger, Julia K and Brown, Andrew},
  journal={Energy Research \& Social Science},
  volume={66},
  pages={101486},
  year={2020},
  publisher={Elsevier},
  doi={10.1016/j.erss.2020.101486}
}

@article{gossling2019sca,
	author = {G{\"o}ssling, Stefan and Choi, Andy and Dekker, Kaely and Metzler, Daniel},
	journal = {Ecological Economics},
	pages = {65--74},
	publisher = {Elsevier},
	title = {The social cost of automobility, cycling and walking in the {E}uropean {U}nion},
	volume = {158},
	year = {2019},
	doi={10.1016/j.ecolecon.2018.12.016}
}

@article{gossling2020cities,
  title={Why cities need to take road space from cars-and how this could be done},
  author={G{\"o}ssling, Stefan},
  journal={Journal of Urban Design},
  volume={25},
  number={4},
  pages={443--448},
  year={2020},
  publisher={Taylor \& Francis},
  doi={10.1080/13574809.2020.1727318}
}

@article{world2022walking,
  title={Walking and cycling: latest evidence to support policy-making and practice},
  author={{WHO}},
  year={2022},
  publisher={World Health Organization. Regional Office for Europe}
}

@article{nieuwenhuijsen2020utp,
  title={Urban and transport planning pathways to carbon neutral, liveable and healthy cities; A review of the current evidence},
  author={Nieuwenhuijsen, Mark J},
  journal={Environment international},
  pages={105661},
  year={2020},
  publisher={Elsevier},
  doi={10.1016/j.envint.2020.105661}
}

@article{branioncalles2020ccr,
	Author = {Michael Branion-Calles and Thomas G\"otschi and Trisalyn Nelson and Esther Anaya-Boig and Ione Avila-Palencia and Alberto Castro and Tom Cole-Hunter and Audrey de Nazelle and Evi Dons and Mailin Gaupp-Berghausen and Regine Gerike and Luc Int Panis and Sonja Kahlmeier and Mark Nieuwenhuijsen and David Rojas-Rueda and Meghan Winters},
	Journal = {Accident Analysis and Prevention},
	Pages = {},
	Title = {{Cyclist crash rates and risk factors in a prospective cohort in seven European cities}},
	Volume = {141},
	Year = {2020},
	doi={10.1016/j.aap.2020.105540}
}

@article{hartmann2020hoa,
  title={How {O}slo achieved zero},
  author={Hartmann, Anders and Abel, Sarah},
  journal={ite journal},
  pages={32--38},
  year={2020},
  publisher={Institute of transportation engineers}
}

@article{marshall2019wcw,
	Author = {Wesley E. Marshall and Nicholas N. Ferenchak},
	Journal = {Journal of Transport \& Health},
	Pages = {285-301},
	Title = {{Why cities with high bicycling rates are safer for all road users}},
	Volume = {13},
	Year = {2019},
	doi={10.1016/j.jth.2019.03.004}
}

@article{bahrololoom2020mis,
	Author = {Sareh Bahrololoom and William Young and David Logan},
	Journal = {Accident Analysis and Prevention},
	Pages = {},
	Title = {{Modelling injury severity of bicyclists in bicycle-car crashes at intersections}},
	Volume = {144},
	Year = {2020},
	doi={10.1016/j.aap.2020.105597}
}

@article{ling2019cmv,
	Author = {Rebecca Ling and Linda Rothman and Marie-Soleil Cloutier and Colin Macarthur and Andrew Howard},
	Journal = {Accident Analysis and Prevention},
	Pages = {},
	Title = {{Cyclist-motor vehicle collisions before and after implementation of cycle tracks in Toronto, Canada}},
	Volume = {135},
	Year = {2020},
	doi={10.1016/j.aap.2019.105360}
}

@article{aldred2018cir,
  title= {{Cycling injury risk in London: A case-control study exploring the impact of cycle volumes, motor vehicle volumes, and road characteristics including speed limits}},
  author={Aldred, Rachel and Goodman, Anna and Gulliver, John and Woodcock, James},
  journal={Accident Analysis \& Prevention},
  volume={117},
  pages={75--84},
  year={2018},
  publisher={Elsevier},
  doi={10.1016/j.aap.2018.03.003}
}

@misc{Redmon2018,
  author = {Redmon, Joseph and Farhadi, Ali},
  booktitle = {arXiv},
  issn = {23318422},
  title = {{YOLOv3: An incremental improvement}},
  year = {2018}
}

@misc{colvilleandersen2017asc,
	author = {Colville-Andersen, Mikael},
	howpublished = {http://www.copenhagenize.com/2017/05/arrogance-of-space-copenhagen-hans.html},
	url={http://www.copenhagenize.com/2017/05/arrogance-of-space-copenhagen-hans.html},
	note = {Copenhagenize},
	title = {Arrogance of Space - {C}openhagen - {H}ans {C}hristian {A}ndersen {B}oulevard},
	year = {2017}}

@book{nacto2014ubd,
	author = {NACTO},
	publisher = {Island Press},
	title = {Urban bikeway design guide},
	year = {2014}}

@article{szell2018cqv,
	author = {Szell, Michael},
	doi = {10.17645/up.v3i1.1209},
	journal = {Urban Planning},
	pages = {1--20},
	title = {Crowdsourced quantification and visualization of urban mobility space inequality},
	volume = {3},
	year = {2018}}

\end{document}